\documentclass{article}
\usepackage{ijcai19}

\usepackage{times}
\usepackage{soul}
\usepackage{url}
\usepackage[draft]{hyperref}
\usepackage[utf8]{inputenc}
\usepackage[small]{caption}
\usepackage{graphicx,tabularx,booktabs}
\usepackage{booktabs}
\usepackage{xspace}
\usepackage{lipsum}
\usepackage{verbatim}
\usepackage{makecell}
\usepackage{mathtools}

\newcommand{\sameAs}{\texttt{owl:sameAs}\xspace}
\newcommand{\sameAsProblem}{``sameAs problem''\xspace}

\title{The sameAs Problem: A Survey on Identity Management in the Web of Data}

\author{
Joe Raad$^1$\and
Nathalie Pernelle$^2$\and
Fatiha Sa{\"i}s$^{2}$\and
Wouter Beek$^1$\And
Frank van Harmelen$^1$\\
\affiliations
$^1$Department of Computer Science, VU University, Amsterdam, The Netherlands\\
$^2$LRI, Paris-Sud University, CNRS 8623, Paris-Saclay University, Orsay, France\\
\emails
\{j.raad, w.g.j.beek, frank.van.harmelen\}@vu.nl, \{nathalie.pernelle, fatiha.sais\}@lri.fr
}

\begin{document}

\maketitle

\begin{abstract}
In a decentralised knowledge representation system such as the Web of Data, it is common and indeed desirable for different knowledge graphs to overlap. Whenever multiple names are used to denote the same thing, \sameAs statements are needed in order to link the data and foster reuse. Whilst the deductive value of such identity statements can be extremely useful in enhancing various knowledge-based systems, incorrect use of identity can have wide-ranging effects in a global knowledge space like the Web of Data. With several works already proven that identity in the Web is broken, this survey investigates the current state of this ``sameAs problem''. An open discussion highlights the main weaknesses suffered by solutions in the literature, and draws open challenges to be faced in the future.
\end{abstract}

\section{Introduction}
\label{sec:introduction}
In the era where the field of Artificial Intelligence (AI) is strongly dominated by Machine Learning, it is sometimes forgotten that the past decade has also seen a major breakthrough in Knowledge Representation (KR). Through the combination of web-technologies and a judicious choice of formal expressivity (description logics which correspond to a decidable 2-variable fragment of first order logic), it has become possible to construct and reason over knowledge graphs of sizes that were not imaginable only few years ago. Nowadays, knowledge graphs of hundreds of millions of statements are routinely deployed by researchers from various fields (e.g. computer science, medicine, humanities), and companies worldwide (e.g. Google, Bing, Facebook). Since these knowledge graphs are mostly developed independently of one another, it is important that different organisations adhere to common principles and standards for encoding and publishing their knowledge. The most adopted set of principles were laid out by Tim Berners-Lee in 2010, and are known as the Linked Open Data (LOD) principles\footnote{\url{https://www.w3.org/DesignIssues/LinkedData.html}}. The idea is by providing simple best practices for creating structured data, publishers can also enrich, access, and benefit from a larger decentralised knowledge graph, known as the Web of Data. 

In such a large and distributed knowledge graph, it is common practice for the same real-world entity to be described in different knowledge graphs. In the absence of a central naming authority in the Web of Data, it is unavoidable for this same real-world entity to be denoted by different names (IRIs, literals, blank nodes). Hence, essential to the coherence of these large and geographically distributed knowledge graphs, publishers are encouraged to link their data. Such interlinking is typically established by asserting that two names denote the same real-world entity. For this purpose, the Web Ontology Language (OWL) introduced in 2004 the \sameAs predicate\footnote{\url{https://www.w3.org/TR/owl-ref\#sameAs}}. For instance, $\langle dbr:Barack\_Obama,$ \sameAs, $wdt:Q76 \rangle$ states that both names from the \textit{DBpedia} and \textit{Wikidata} knowledge graphs refer to the same entity. With its strict logical semantics, this statement indicates that every property asserted to one name will also be inferred to the other. Hence, allowing both names to be used interchangeably in all contexts. 

While such inferences can be extremely useful in enhancing a number of knowledge-based systems (e.g. providing more coverage and context for search engines, virtual assistants and recommendation systems), incorrect use of identity can have wide-ranging effects in a global knowledge space like the Web of Data. In fact, a number of studies over the years have already shown that identity is misused, estimating the number of existing erroneous \sameAs in the Web of Data to be between 2.8\% \cite{hogan2012scalable} and 20\% \cite{halpin2010owl}.
In addition, by exploiting the semantics of \sameAs and computing the transitive closure of over half a billion statements, \cite{raad2018detecting} showed the effects of such identity misuse in practice. Specifically, it shows that whilst in some cases identity misuse results in the false equivalence of semantically close entities (e.g. \textit{Barack Obama} and the \textit{Obama administration}), other cases have resulted in the false equivalence of over 177K names referring to a number of different countries, cities and people. With such findings leaving many uncertainties over the quality and usability of the Web of Data in its current state, a proper approach towards the handling of identity links is required in order to make the Web of Data succeed as an integrated knowledge space.\\


This survey provides the first overview of existing approaches to this widely recognised identity problem in the Web of Data, known as the \sameAsProblem \cite{halpin2010owl}. It describes these different solutions, discusses their strengths and limitations, and formulates open challenges. This survey does not cover related but distinct research topics such as entity resolution \cite{ferrara2013data,nentwig2017survey} and ontology alignment \cite{euzenat2007ontology}, that focus on techniques  and frameworks  for establishing \sameAs links. In addition, this survey does not address the historically significant distinction between locating an electronic document with a URL and denoting an RDF resource with an IRI, known as the problem of `Sense and Reference' \cite{halpin2011sense}. The rest of this paper is structured as follows. Section \ref{sec:problems} gives an overview of the various aspects of the identity problem. Section \ref{sec:alternative-identity-links-soa} presents existing alternative identity relations to \sameAs. Section \ref{sec:identity-services-soa} gives an overview on proposed strategies and services for managing identity in the Web. Section \ref{sec:erroneous-identity-soa} covers existing approaches for the detection of erroneous identity links, and Section \ref{sec:discussion} concludes and formulates open challenges.

\section{Identity Overview}\label{sec:problems}
Identity is an old and thorny topic. Classically speaking, entities that are identical are considered to share the same properties. With $N$ denoting the set of all names, and $\Psi$ the set of all properties, this `Indiscernibility of Identicals' (\ref{eq:1}) is attributed to Leibniz and its converse, the `Identity of Indiscernibles' (\ref{eq:2}) states that entities that share the same properties are identical. That identity is reflexive, symmetrical and transitive also follows from Leibniz's Law.
\begin{gather}
a=b \rightarrow
(\forall_{\psi\in\Psi})(\psi(a)=\psi(b))\label{eq:1}\\
(\forall_{\psi\in\Psi})(\psi(a)=\psi(b)) \rightarrow a=b\label{eq:2}
\end{gather}
 This identity relation induces a partitioning of $N$ into a collection of non-empty and mutually disjoint \textit{equivalence classes} $N_k \subseteq N$. From the premises $\psi(a)$, and $a, b \in N_k$, it follows that $\psi(b)$ is also the case. In fact, this deduction is central to the Web of Data as it allows complementary descriptions of the same resource to be maintained locally, yet interchanged globally, merely by interlinking the names that are used in those respective descriptions. However, there are also problems with it, and -- consequently -- criticisms have been levelled against it. These problems are not new, neither specific to the Web of Data, as they are present in all KR systems \cite{GrantS95,Nguyen:2007}. However, the problems are specifically pressing in the Web of Data due to its unprecedented size, the heterogeneity of its content and users, and the absence of a central naming authority. This section briefly presents some of the well-known issues with this notion of identity.

\subsection{Philosophical Problems}\label{sec:problem-1}
From a philosophical point of view, we present the two major issues with this notion of identity. Firstly, identity over time poses problems, since a ship\footnote{Reference to the ship of Theseus or Theseus's paradox} may still be considered the same ship, even though some, or even all, of its original components (i.e. properties) have been replaced by new ones \cite{lewis1986plurality}. In addition, identity is context-dependent \cite{Geach1967}, allowing two medicines, having the same chemical structure, to be considered the same in a medical context, but to be considered different in other contexts (e.g. because they are produced by different companies). These issues in the classical identity definition have led to various philosophical theories, such as the distinction between accidental properties (traits that could be taken away from an object without making it a different thing), and essential properties (core elements needed for a thing to be the thing that it is) \cite{kripke1972naming}.

\subsection{Practical Problems}\label{sec:problem-2}
Given that this problematic notion of identity is also standardised as part of the Web Ontology Language, it is normal to encounter these issues in Web applications. In fact, and due to the Open World Assumption and the continuous increase of $\Psi$, identity statements in the Web of Data are even more controversial. Firstly, unless two things are explicitly said to be different (e.g. using \texttt{owl:differentFrom}), the absence of an identity statement between them does not mean that they are not identical. Compared to the 558M \sameAs present in a 2015's crawl of the Web of Data \cite{Fernandez2017}, this type of statements is barely present in the Web of Data, with only 3.6K \texttt{owl:differentFrom} statements existing at that time in the same dataset. In addition, most \sameAs links are generated by heuristic entity resolution techniques, that employ practical strategies which are not guaranteed to be accurate. For instance, the precision of such tools ranged between 67\% and 86\% in the 2017 and 2018 Ontology Alignment Evaluation Initiative (OAEI)\footnote{\url{http://oaei.ontologymatching.org/2018/results/conference/index.html}}.
Finally, studies have shown that modellers have different opinions about whether two objects are the same or not. For instance in \cite{halpin2010owl}, three KR experts were asked to judge 250 \sameAs links collected from the Web. The evaluation shows high disagreements, with one judge confirming the correctness of only 73 \sameAs statements, whilst the two other experts judging up to 132 and 181 links as true. While in some cases this may be due to differences in modelling competence, there is also the problem that two modellers may consider different parts of the same knowledge graph within different contexts.

\section{Alternative Identity Links}\label{sec:alternative-identity-links-soa}
Given these presented problems in \sameAs, a number of vocabularies and approaches have proposed alternative identity relations. This section presents the most deployed alternatives and gives an overview of their usage in Table \ref{table:overview-alternative-links}.

\subsection{Weak-Identity and Similarity Predicates}\label{sec:weak-identity}
\noindent {\bf SKOS predicates.} Introduced as lighter alternatives for \sameAs with \texttt{skos:closeMatch} indicating that ``two concepts are sufficiently similar that they can be used interchangeably in some applications", and \texttt{skos:exactMatch} indicating ``a high degree of confidence that the concepts can be used interchangeably across a wide range of applications". \\[1.10ex]
\noindent {\bf wdt:P2888.} In Wikidata the exact match predicate (P2888), declared as equivalent to \texttt{skos:exactMatch}, is deployed for linking concepts. \\[1.10ex]
\noindent {\bf umbel:isLike.} This symmetrical relation was introduced by the UMBEL vocabulary to ``assert an associative link between similar individuals who may or may not be identical, but are believed to be so". \\[1.10ex]
\noindent {\bf Similarity Ontology.} \cite{halpin2010owl} introduced eight new predicates hierarchically represented with existing RDFS, OWL and SKOS predicates. Each predicate in this ontology is also characterised by reflexivity, transitivity and symmetry. The most specific predicate in this ontology is \sameAs, and the most general ones are \texttt{so:claimsRelated} and \texttt{so:claimsSimilar}.

\begin{table}
	\caption{Overview of the usage of alternative identity links, based on a 2015 crawl of the Web of Data, and Wikidata for \textit{wdt:P2888}.}
	\centering
	\label{table:overview-alternative-links}
	\scriptsize{
	\begin{tabular}{| l | l | l |}
		\hline
		\textbf{Property}        & \textbf{Unique Triples} & \textbf{Unique Names}\\
		\hline
		\hline
		\texttt{owl:sameAs} & 558,943,116     & 179,739,567      \\
		\hline
		\texttt{skos:exactMatch} & 566,137            & 1,087,866       \\
		\hline
		\texttt{umbel:isLike}    & 461,054            & 478,474         \\
		\hline
		\texttt{skos:closeMatch} & 371,011            & 647,230         \\
		\hline
		\texttt{wdt:P2888}    & 356,648           & 696,535     \\
		\hline
	\end{tabular}
	}
\end{table}

\subsection{Contextual Identity} \label{sec:contextual-identity}
The standardised semantics of \sameAs can be thought of as instigating an implicit context that is characterised by all (possible) properties to have the same values for the linked names. Weaker types of identity can be expressed by considering a subset of properties with respect to which two resources can be considered the same. At the moment, the way of encoding contexts on the Web is largely ad hoc, as contexts are often embedded in application programs, or implied by community agreement. The issue of deploying contexts in KR systems has been extensively studied in AI \cite{guha1991contexts}. In the Web of Data, explicit representation of context has been a topic of discussion since its early days \cite{bouquet2003c}, where the variety and volume of the web poses a new set of challenges than the ones encountered in previous AI systems. This section presents approaches focusing on the specific issue of representing contextual identity in the Web.

In \cite{beek2016contextualised}, a context $\Pi$ is defined as a subset of all properties $\Psi$ which are necessary and sufficient to determine indiscernibility and hence identity: 
\begin{gather}
    a =_\Pi b \rightarrow
(\forall_{\pi\in\Pi})(\pi(a)=\pi(b)) \label{eq:3}\\
(\forall_{\pi\in\Pi})(\pi(a)=\pi(b)) \rightarrow  a =_\Pi b\label{eq:4}
\end{gather}
Looking back to the example in Section 2.1, two medicines with the same chemical structure, but produced by different companies, are identical in the context where the property $\pi_i$ specifying the medicine's commercial supplier is discarded (i.e. $\pi_i \not\in \Pi$). In \cite{raad2017detection}, this notion of contextual identity is encoded in RDF, and the definition of a context is extended to a sub-graph of the domain ontology called a \textit{global context}. Specifically, a global context $\mathcal{G}$ is composed of a subset of classes $C_\mathcal{G}$ and properties $P_\mathcal{G}$ of an ontology $\mathcal{O}$, and a set of axioms which are limited to constraints on property domains and ranges. These axioms allow the parameterization of the identity criteria with respect to each class of the ontology. For instance, allowing to express that two medicines are considered identical if they have the same quantity of elements of type $c_1$, whilst disregarding the quantity of its other elements. The identity relation between two class instances in a global context is based on the notion of graph isomorphism of their descriptions, where an approach is proposed for automatically detecting these global contexts.

With both these approaches unclear about the treatment of properties $p$ that do not belong to the identity context (i.e. $p \notin \Pi$ or $p \notin P_\mathcal{G}$), a richer definition of context was proposed by \cite{idrissou2017my}. It defines a context by two sets of properties, $\Gamma$ for indiscernibility and $\Lambda$ for propagation:
\begin{gather}
    a =_{(\Gamma,\Lambda)} b \rightarrow
(\forall_{\gamma\in\Gamma})(\gamma(a)=\gamma(b))\label{eq:5}\\
(\forall_{\gamma\in\Gamma})(\gamma(a)=\gamma(b)) \rightarrow  a =_{(\Gamma,\Lambda)} b\label{eq:6}\\
a =_{(\Gamma,\Lambda)} b \rightarrow (\forall_{\lambda\in\Lambda})(\lambda(a)=\lambda(b))\label{eq:7}
\end{gather}
Principles (\ref{eq:5}) and (\ref{eq:6}) refers to the same notion of contextual identity defined in \cite{beek2016contextualised}, whilst (\ref{eq:7}) defines the notion of \textit{contextualised propagation}. Note that unlike $\Gamma$, indiscernibility in $\Lambda$ does not determine identity. For instance, in a scientific context, two medicines sharing the same chemical structure $\gamma_1$ is enough to consider them identical, and infer that they share the same purpose $\lambda_1$. However, two medicines with the same $\lambda_1$ do not necessarily share the same $\gamma_1$. This approach extends a previous approach by \cite{batchelor2014scientific}, mainly in the way of parametrizing the propagation context $\Lambda$, and the way these contextual identity links are encoded in RDF (on the triples level instead of the graphs level).

\section{Identity Management Services}\label{sec:identity-services-soa}
Instead of proposing alternative identity relations for limiting the misuse of \sameAs, other approaches have proposed services for managing identity in the Web of Data. These services share the common goal of helping users or applications to identify names referring to the same real world entity, and distinguish between similar labels referring to different real world entities. For instance, in order to avoid using a name referring to the \textit{river of Niger}, while intending in using one referring to the \textit{country of Niger}, one could benefit from such services for re-using an existing universal identifier that unambiguously refers to a specific real-world entity (e.g. \textit{river of Niger}). Such type of services have a more centralised vision for identity management in the Web of Data, in which each real-world entity is referenced by a single centralised name. On the other hand, one can make use of other types of services that provide centralised access to identity statements that are published in a decentralised way. Such \textit{identity observatories} allow Web consumers to make an informed decision regarding the quality of identity statements they encounter. Such services can also play an important role in enabling large scale identity analysis in the Web \cite{beek2018sameas}, implementing and optimising linked data queries in the presence of co-reference \cite{schlegel2014balloon}, and detecting erroneous identity statements \cite{Melo13}. This section gives an overview of existing identity services.

\subsection{Centralised Identity Management}\label{sec:centralised}
In the early days of the Web, it was originally conceived that resource identifiers would fall into two classes: locators (URLs) to identify resources by their locations, and names (URNs) for assigning location-independent, globally unique, and persistent identifiers \cite{Mealling1999}. With URNs, each identifier has a defined namespace that is registered with the Internet Assigned Numbers Authority (IANA). For instance, \textit{urn:isbn:0451450523} is a URN that identifies the book ``The Last Unicorn'', using the ISBN registered namespace. Because of the lack of a well-defined resolution mechanism, and the organisational hurdle of requiring registration with IANA, URNs are hardly used (total of 47K URNs in a 2015 crawl of the Web of Data, with only 73 registered\footnote{https://iana.org/assignments/urn-namespaces} URN namespaces with IANA at the time of writing). Since 2005, the use of the terms URNs and URLs has been deprecated in favour of the terms URI which encompasses both, and IRI that extends the URI character set. A more recent centrally managed naming service was proposed by \cite{bouquet2007okkam}. This public entity name service named Okkam\footnote{as a variation of Occam's razor}, intends to establish a global digital space for publishing and managing information about entities, with the idea of encouraging people to reuse existing names instead of creating new ones. Every entity is uniquely identified with an unambiguous universal name known as an OKKAM ID, and is matched to a set of existing names (e.g. DBpedia and Wikidata names). In addition, for each OKKAM entity, a set of attributes are collected and stored in the service for the purpose of finding and distinguishing entities from another. However, this public service\footnote{hosted at \url{http://okkam.org}} is no longer maintained, with no information on the number of existing entities and links. 
 
\subsection{Identity Observatories}\label{sec:observatories}
In recent years, identity observatories have gained more popularity. These web services, compared in Table \ref{table:overview-identity-services}, allow users to find for a given name, the list of names that belong to the same equivalence class. Whilst in recent services, these equivalence classes are based solely on the transitive closure of \sameAs statements, the Consistent Reference Service \footnote{hosted at \url{http://sameas.org}} \cite{glaser2009managing} incorporates a mix of identity and similarity relationships (such as \sameAs, \texttt{umbel:isLike}, and the SKOS predicates), harvested from multiple RDF dumps and SPARQL endpoints. 
On the other hand, the LODsyndesis\footnote{hosted at \url{http://www.ics.forth.gr/isl/LODsyndesis}} co-reference service is based on the transitive closure of solely \sameAs statements harvested from existing data dumps (e.g. \texttt{datahub.io}, subsets of \texttt{DBpedia} and \texttt{Wikidata}). Finally, the recent co-reference service\footnote{hosted at \url{http://sameas.cc}} introduced by \cite{beek2018sameas} provides the largest collection of \sameAs with their equivalence closure collected from a 2015 crawl of the Web of Data.
\begin{table}
	\caption{Overview of Existing Identity Observatories}
	\centering
	\label{table:overview-identity-services}
	\scriptsize{
	\begin{tabular}{| l | l | l | l |}
		\hline
		\textbf{}  
		& \textbf{sameas.org} 
		& \textbf{LODsyndesis}
		& \textbf{sameas.cc} \\
		\hline
		\hline
		\textit{\# Names} 
		& 203,953,936      
		& 65,315,931 
		& 179,739,567   \\
		\hline
		\textit{\# Statements} 
		& 346,425,685     
		& 44,028,829 
		& 558,943,116 \\
		\hline
		\textit{\# owl:sameAs} 
		& Unknown         
		& 44,028,829  
		& 558,943,116 \\
		\hline
		\textit{\# Partitions} 
		& 62,591,808    
		& 24,076,816
		& 48,999,148  \\
		\hline
		\textit{\# Eq. Classes} 
		& Unknown   
		& 24,076,816
		& 48,999,148  \\
		\hline
	\end{tabular}
			}
\end{table}

\section{Erroneous Identity Links Detection}\label{sec:erroneous-identity-soa}
Finally, an important aspect of limiting the \sameAsProblem is the detection of incorrectly asserted identity links. In order to detect such incorrect links, various kinds of information may be exploited: RDF triples related to the linked resources, domain knowledge that is described in the ontology or that is obtained from experts, or different network metrics. This section presents existing approaches, classified into three -- possibly overlapping -- categories. Table \ref{table:detection} provides an overview of these approaches. 

\begin{table*}[ht]
\centering
\caption{Overview of erroneous identity links detection approaches, stating their type, requirements, the dataset on which the experiments were conducted, and the reported results. Transparency indicates whether the dataset (D), the tool (T), and the results (R) were made available.}
\label{table:detection}
\resizebox{\linewidth}{!}{%
\begin{tabular}{|c|c|c|c|c|c|c|}
\hline
\textbf{Approach}          
& \textbf{Type of Approach}       & \textbf{Requirements}      
& \textbf{Evaluated Data} 
& \textbf{Results} 
& \textbf{\begin{tabular}[c]{@{}c@{}}Transparency\end{tabular} } 

\\ \hline
\cite{cudre2009idmesh}   
&  \thead{Inconsistency-based}      
&  \thead{- Source Trustworthiness \\- Presence of owl:differentFrom}   
& \thead{Synthetic graph of \\8K entities and 24K links}          
& \thead{75\% to 90\% accuracy} 
& -

\\ \hline
\cite{hogan2012scalable}   
&  \thead{Inconsistency-based}      
&  \thead{Ontology Axioms}   
& \thead{3.77M \sameAs from a\\2010 crawl of 3.9M Web documents}   
& \thead{85\% precision, 40\% recall (only\\280 inconsistent classes out of 2.8M)} 
& -

\\ \hline
\cite{papaleo2014logical}   
&  \thead{Inconsistency-based\\ and Content-based}     
&  \thead{- Ontology Axioms\\- Ontology Mappings} 
& \thead{344 \sameAs produced by\\3 different linking tools (OAEI 2010)}   
& \thead{37\% to 88\% precision, 75\% to 100\%\\recall (depending on the dataset)} 
& D

\\ \hline
\cite{Melo13}   
&  \thead{Inconsistency-based}      
&  \thead{UNA}   
& \thead{BTC2011: 3.4M \sameAs and\\sameAs.org: 22.4M \sameAs}   
& \thead{no precision or\\recall evaluation} 
& D

\\ \hline
\cite{valdestilhas2017cedal}   
&  \thead{Inconsistency-based}      
&  \thead{UNA}   
& \thead{LinkLion: 19.2M \sameAs}   
& \thead{no precision or\\ recall evaluation} 
& D, T, R


\\ \hline
\cite{paulheim2014identifying}   
&  \thead{Content-based\\(outlier detection)}      
&  \thead{-}   
& \thead{Peel-DBpedia: 2K \sameAs\\DBTropes-DBpedia: 4.2K \sameAs}   
& \thead{- 58\% to 80\% AUC\\- 50\% F1-measure} 
& D, T   

\\ \hline
\cite{cuzzola2015filtering}   
&  \thead{Content-based\\(natural language analysis)}  
&  \thead{Textual Description\\for each resource} 
& \thead{sameas.org: 411 \sameAs\\(from 7K collected ones before cleansing)}   
& \thead{93\% precision\\75\% recall} 
& -

\\ \hline
\cite{gueret2012assessing} 
& \thead{Network Metrics\\(local network)}  
& \thead{-}     
& \thead{SILK framework: 100 \sameAs}
& \thead{49\% precision\\68\% recall}    
& D, T, R

\\ \hline
\cite{raad2018detecting} 
& \thead{Network Metrics\\(identity network)}  
& \thead{-}     
& \thead{558.9M \sameAs from a \\2015 crawl of the Web of Data}
& \thead{93\% recall, 40\% to 73\% precision\\(depending on the eq. class size)}    
& D, T, R

\\ \hline
\end{tabular}%
}
\end{table*}

\subsection{Inconsistency-based Detection Approaches}\label{sec:inconsistency-based}
This category of approaches hypothesises that \sameAs assertions leading to logical inconsistencies must be wrong.

\subsubsection{Conflicting owl:sameAs and owl:differentFrom}
In \cite{cudre2009idmesh}, these logical inconsistencies are restricted to conflicting \sameAs and \texttt{owl:differentFrom} statements. These conflicts are detected based on a graph-based constraint satisfaction problem that exploits the symmetry and transitivity of \sameAs statements. These detected conflicts are resolved based on the iteratively refined trustworthiness of the sources declaring the statements (i.e. hypothesises that links published by trusted sources are more likely to be correct). The approach shows high accuracy (75 to 90\%), with the evaluation only conducted on synthetic data involving 24K links.

\subsubsection{Ontology Axioms Violation}
In \cite{hogan2012scalable}, logical inconsistencies are detected after transitive closure, by exploiting ten OWL 2 RL/RDF rules expressing the semantics of axioms such as \textit{differentFrom, AsymmetricProperty}. When entities causing inconsistencies are detected, they are separated into different seed equivalence classes, then the remaining entities are assigned into one of these seed classes based on their minimum distance in the equivalence class. The approach manages to detect inconsistencies in 280 out of the 2.8M equivalence classes that resulted from the closure 3.7M \sameAs. The approach shows high precision (85\%) and lower recall (40\%). These results also show that consistency does not imply correctness, with 60\% of the pairs manually evaluated as being different still belong in the same -- now consistent -- equivalence class. In \cite{papaleo2014logical}, the authors exploit class disjointness, (inverse) functional properties and locally complete properties\footnote{multi-valued properties where its information is complete when it is present (e.g., the authors of a certain publication).} for detecting inconsistencies. Firstly the approach builds a contextual graph of a specified depth describing the two involved resources in an identity link, then applies a Unit-resolution inference rule until saturation for detecting inconsistencies within these graphs. The approach was evaluated on three datasets with a total of 344 \sameAs, showing low precision in two (37\% and 42\%) and an 88\% precision in the third, with a recall between 75\% and 100\%.

\subsubsection{Unique Name Assumption Violation}
In \cite{Melo13} and \cite{valdestilhas2017cedal}, inconsistencies are detected by presuming that knowledge graphs preserve the Unique Name Assumption (UNA), and that violations of the UNA are indicative of erroneous identity links. The UNA indicates that two names in the same knowledge graph, do not refer to the same real-world entity. Experiments show that both approaches are scalable (tested on 26M and 19M \sameAs respectively). However, the precision, recall and accuracy of both approaches have not been evaluated. Interestingly, \cite{Melo13} claims that most of the UNA violations stem from incorrect identity links, not from inadvertent duplicates. Whilst in the analysis of a sample of 100 errors, the authors of \cite{valdestilhas2017cedal} show that 90\% of the errors stem from duplications within the dataset, instead of referring to two different real world entities. 

\subsection{Content-based Approaches}
\label{sec:content-based}
This category of approaches exploits the descriptions associated to each name for evaluating the correctness of an identity link. 
In \cite{paulheim2014identifying}, the author hypothesises that correct identity links follow certain patterns, with ones violating those patterns being probably erroneous. The approach represents each identity link as a feature vector, and tests six different methods for detecting outliers (e.g. one-class support vector machines). The evaluation conducted on two different datasets (2K and 4K \sameAs each), shows a maximum F1-measure of 54\%, that varies between each dataset. Finally, the authors in \cite{cuzzola2015filtering} used DBpedia categories for calculating a similarity score of the textual descriptions associated to (claimed) identical pairs. The approach was tested on 411 \sameAs links, with the evaluation suggesting a precision between 86\% and 93\%, and a recall between 75\% and 79\%.

\subsection{Network-based Approaches}
Finally, a last category of approaches used network metrics for evaluating the quality of \sameAs links. Whilst in \cite{raad2018detecting} the exploited (identity) network solely contains \sameAs statements, in \cite{gueret2012assessing} the (local) network considers all properties and names related to the two names linked by an \sameAs. Specifically, this approach aims at measuring the impact that a given \sameAs has on this local network, using three classic network metrics (clustering coefficient, betweenness centrality, and degree) and two Linked Data-specific ones (description richness and \sameAs chains). For instance in the latter, it hypothesises that a correct \sameAs will contribute in closing an open \sameAs chain. The evaluation was conducted on a set of 100 links, and shows a 49\% precision, and 68\% recall. In \cite{raad2018detecting}, the approach hypothesises that the more densely a group of names is interlinked, the higher the likelihood of those names to be identical. The approach firstly partitions the identity network into different connected components and then detects the community structure in each of these components. Finally, it assigns an error degree to each \sameAs based on the density of the community(ies) in which the two interlinked names belong and the reciprocity of the link. The evaluation was conducted on the \textit{sameas.cc} dataset, and shows a precision between 40\% and 73\% and a recall of 93\%.

\section{Conclusion \& Discussion}\label{sec:discussion}

This survey has presented the first overview in the ongoing process of limiting the excessive and incorrect use of identity links in the Web of Data. We now present the current situation, and set out directions for future work.\\[1.10ex]
\textbf{Existing identity links lack semantics.} In Section \ref{sec:weak-identity}, several alternative identity predicates were presented. A big downside of these alternatives is their lack of formal semantics. For instance, in \texttt{skos:exactMatch} whether a degree of confidence is high (enough) is subjective, and the meaning of this relation even changes over time, because information is always evolving over time. Also, some proposed alternative properties do not denote equivalence relations, which means that they are of limited use in reasoning and linking. Another downside of these approaches is that they require data publishers to change their modelling practice. A lot of momentum is needed in order to create new knowledge graphs, or to change existing ones in order to make use of these alternative properties. As a result, most of these proposals lack uptake and are only used in a handful of datasets (see Table \ref{table:overview-alternative-links}).\\[1.10ex]
\textbf{Contextual identity requires further investigation.} In Section \ref{sec:contextual-identity}, different proposals for context-dependent semantics of identity were presented. These approaches have the benefit that they do not require existing modelling practices to be changed since the same property (i.e., \sameAs) can be used. An exception to this are approaches that require contexts to be modelled by hand. However, contextual semantics has not yet been widely implemented in Linked Data tools, e.g., reasoners, linked data browsers, and faces potential impediments for uptake. In fact, the exact impact of contextual identity on entailment has not been sufficiently investigated. Finally, the use of identity assertions for the purpose of interlinking may be somewhat hampered by contextual semantics approaches. With the traditional semantics of \sameAs, linked descriptions can always be shared, but with contextual semantics such descriptions can only be shared if they are asserted in compatible contexts.\\[1.10ex]
\textbf{Centralised naming authorities will be of limited use.} Centralised naming authorities, presented in Section \ref{sec:centralised}, play an important role in facilitating the understanding and re-use of names. However, although they might see limited uptake within some dedicated domains, centralised identity management becomes more difficult and error prone when operating at a larger scale. In addition, the idea of having to go through an authority in order to use a new name somewhat goes against the philosophy of the ad hoc nature of the Web, where ``anybody is able to say anything about anything''.\\[1.10ex]
\textbf{Identity Observatories must be used more broadly.} Even though several identity observatories exist (Section \ref{sec:observatories}), they are not commonly used in Web applications today. This is probably due to the following limitations which these services suffer from, in their current status and architecture.

\textit{- Semantic Interpretability.} The `equivalence classes' in \textit{sameas.org} are the result of the transitive closure of a mix of identity relations with different semantics. Since this service does not keep the original predicates, the semantics of the closure that is calculated is unclear (e.g. can not be used by a DL reasoner for inferring new facts).

\textit{- Coverage.} With the number of statements in \textit{LODsyndesis} being an order of magnitude smaller than other observatories, this service may see limited use in certain applications.

\textit{- Up-to-date support.} With \textit{sameas.cc} being based on a 2015 crawl of the Web, such service may see limited uptake in applications which require more recent information. 

We believe that such services will see uptake over time, since they make it possible to use some of the benefits of linking to other knowledge graphs, while at the same time giving the client some control as to which knowledge graphs to link to (and which ones not to link to).\\[1.10ex]
\textbf{Hybrid error detection approaches are required.} Finally, it has now been broadly acknowledged that erroneous identity statements are present in the Web of Data, and that additional effort is needed in order to detect them. In Section \ref{sec:erroneous-identity-soa}, we have seen that there are several promising approaches for the (semi-) automatic detection of erroneous identity links. However, all existing approaches have made some trade-off, either having less precision, having less recall, or being less scalable. Specifically, experiments in \cite{hogan2012scalable} showed that the Web of Data lack from ontological axioms and assertions that are strong enough for deriving inconsistencies. Hence, suggesting that axiom violation-based approaches will mainly have a lower recall. Experiments based on the UNA violation have showed contradicting results, leaving many uncertainties on the effectiveness of the UNA assumption for the task of detecting erroneous links. Existing content-based approaches have showed promising results, but still requires further investigation in terms of their scalability, and whether sufficient textual descriptions in the Web of Data are indeed available. Finally, network-based approaches have also showed promising results in terms of recall and scalability, but existing experiments showed lower precision.\\[1.10ex]

Future research should focus on combining some of these existing approaches in novel ways, potentially combining some of the strengths of these various approaches into one (hybrid) approach. Such an approach should be feasible over the whole Web, where scalability is not the only challenge, but also where certain assumptions on the constant changing data can not be presumed. For instance, in the Web not all names have textual descriptions, many knowledge graphs do not include vocabulary mappings, or lack semantically rich assertions for deriving inconsistencies. In addition, future research should focus on providing more transparency for allowing other approaches to compare, and hopefully improve, their results. Table \ref{table:detection} shows that only three approaches provide fully reproducible results. Finally, compared to the amount of research invested in entity linking \cite{shen2015entity} and ontology matching \cite{ferrara2013data}, this area is clearly lacking uptake. While in some cases this may be due to various technical challenges (e.g. resulted from the absence of a manually annotated benchmark designed for this task), there is also the aspect that the number and actual effects of these erroneous statements in practice were still unknown, until recently \cite{raad2018detecting}.\\[1.10ex]
\indent With this overview of the current state of the \sameAsProblem, we hope that this survey can lead to the emergence of more efficient approaches and systems for representing contextual identity and investigating its impact at scale, accessing explicit and implicit identity assertions in the Web, and detecting the erroneous ones.

\clearpage
\bibliographystyle{named}
\bibliography{ijcai19}

\end{document}